\documentclass{webofc}
\usepackage{subfig}
\usepackage{graphicx}
\usepackage{booktabs} % for professional tables

\usepackage{amsmath}
\usepackage{amssymb}
\usepackage{mathtools}
\usepackage{amsthm}

\usepackage[capitalize,noabbrev]{cleveref}

\usepackage[varg]{txfonts}   % Web of Conferences font
\usepackage{hyperref}
\usepackage{url}
\hypersetup{colorlinks=true,citecolor=blue,urlcolor=blue,linkcolor=blue}
\begin{document}
\title{Physics and Computing Performance of the EggNet Tracking Pipeline}

\author{\firstname{Jay} \lastname{Chan}\inst{1}\fnsep\thanks{\email{jaychan@lbl.gov}} \and
        \firstname{Brandon} \lastname{Wang}\inst{1,2} \and
        \firstname{Paolo} \lastname{Calafiura}\inst{1}
        % etc.
}

\institute{Scientific Data Division, Lawrence Berkeley National Laboratory, Berkeley, CA 94720, USA
\and
           Department of Computer Science, University of California, Berkeley, CA 94720, USA
          }

\abstract{Particle track reconstruction is traditionally computationally challenging due to the combinatorial nature of the tracking algorithms employed. Recent developments have focused on novel algorithms with graph neural networks (GNNs), which can improve scalability. While most of these GNN-based methods require an input graph to be constructed before performing message passing, a one-shot approach called EggNet that directly takes detector spacepoints as inputs and iteratively apply graph attention networks with an evolving graph structure has been proposed. The graphs are gradually updated to improve the edge efficiency and purity, thus providing a better model performance. In this work, we evaluate the physics and computing performance of the EggNet tracking pipeline on the full TrackML dataset. We also explore different techniques to reduce constraints on computation memory and computing time.
}
\maketitle
\section{Introduction}
Particle track reconstruction is a crucial task in particle experiments. Traditionally, particle tracks are reconstructed with the Combinatorial Kalman Filter algorithm \cite{RevModPhys.82.1419, ATLAS:2017kyn, CMS:2014pgm}. The combinatorial nature of this algorithm leads to a large computational cost and poor scalability (worse than quadratically). To improve the computational cost for future experiments, graph neural network (GNN) has emerged as a promising approach for track reconstruction \cite{ExaTrkX:2021abe, Biscarat:2021dlj, CHEP2022, Lieret:2023aqg} and has been demonstrated with the ability to achieve linear scaling with the event size \cite{ExaTrkX:2021abe, Lazar:2022ixi}.

While promising, most of the existing GNN-based tracking algorithms \cite{ExaTrkX:2021abe, Biscarat:2021dlj, CHEP2022, Lieret:2023aqg} consist of a graph construction step before GNN. The quality of the constructed graphs thus significantly impacts the performance of the subsequent steps in the GNN-based tracking pipelines. For example, graph construction inefficiency can lead to information loss during the GNN message passing step, where information is shared between nodes in a graph along the edges that connect them, while low purity in the constructed graphs (due to fake edges) results in noisy information during the message passing. Both cases affect the performance of GNN.

In \cite{Calafiura:2024qhv}, a new tracking method was proposed to embed the graph construction into an Evolving Graph-based Graph Attention Network (EggNet) architecture. The algorithm takes point clouds as inputs and iteratively constructs graphs based on the updated embedding. The message passing in the EggNet is then based on the updated graph in each iteration. This method continuously improves the constructed graph efficiency in each iteration and thus provides better effectiveness of message passing. A preliminary test with EggNet \cite{Calafiura:2024qhv} on a reduced TrackML dataset \cite{Amrouche:2019wmx, Amrouche:2021nbs} has shown promising track performance compared the existing track reconstruction algorithms.

In this work, we perform training and inference of EggNet on the full TrackML dataset and evaluate the physics and computing performance. In addition, to reduce the computational cost during training, we introduce a strategy to train the EggNet model on segmented subgraphs and compare the performance with this segmented graph training strategy.

% \section{Related Work}

\section{EggNet}

\begin{figure*}[ht]
% \vskip 0.2in
\begin{center}
\centerline{\includegraphics[width=\textwidth]{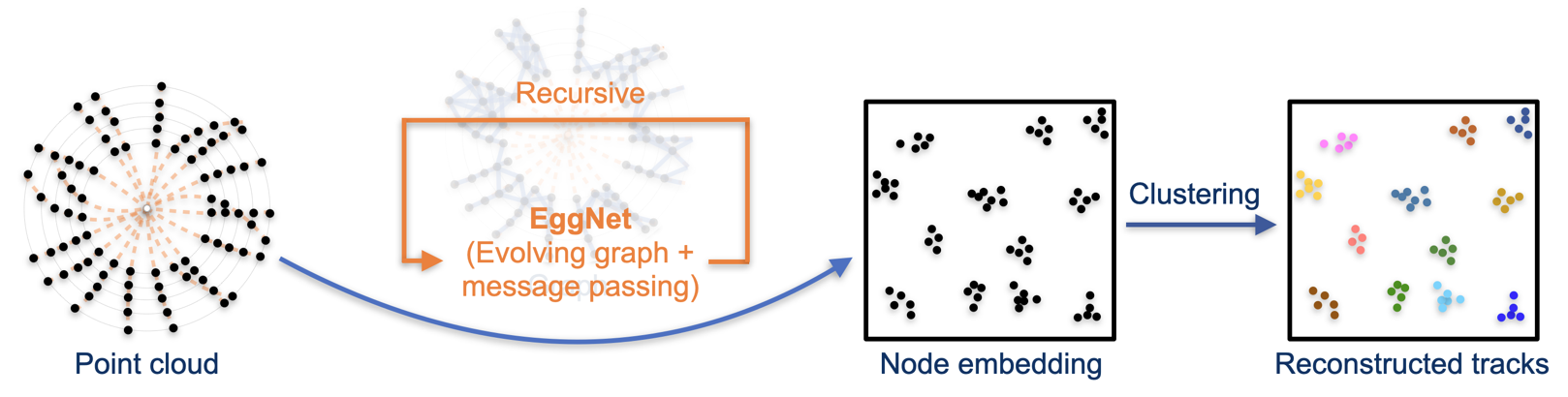}}
\caption{The EggNet tracking pipeline. Starting with a point cloud, where each point corresponds to a spacepoint, we run an EggNet model that iteratively constructs a graph and performs message passing. EggNet outputs the embedding for each node, and the track candidates are extracted from the DBSCAN clusters that are obtained in the node embedding space.}
\label{fig:pipeline}
\end{center}
% \vskip -0.2in
\end{figure*}

As described in details in \cite{Calafiura:2024qhv}, the EggNet tracking pipeline consists of an EggNet step and a clustering step. \cref{fig:pipeline} shows the whole workflow of the pipeline. The EggNet takes a set of spacepoints (a point cloud) as inputs and outputs the node embedding for each node. The nodes are then clustered in the node embedding space with Density-Based Spatial Clustering of Applications with Noise (DBSCAN) \cite{10.5555/3001460.3001507} and assigned to different track candidates.

\begin{figure*}[htbp]
\vskip 0.2in
\begin{center}
\includegraphics[width=0.54\textwidth]{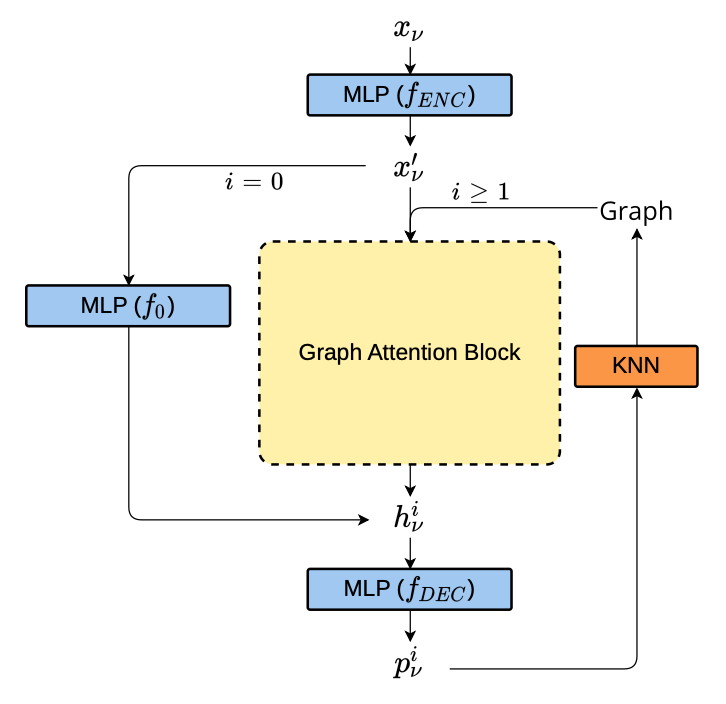}
\includegraphics[width=0.44\textwidth]{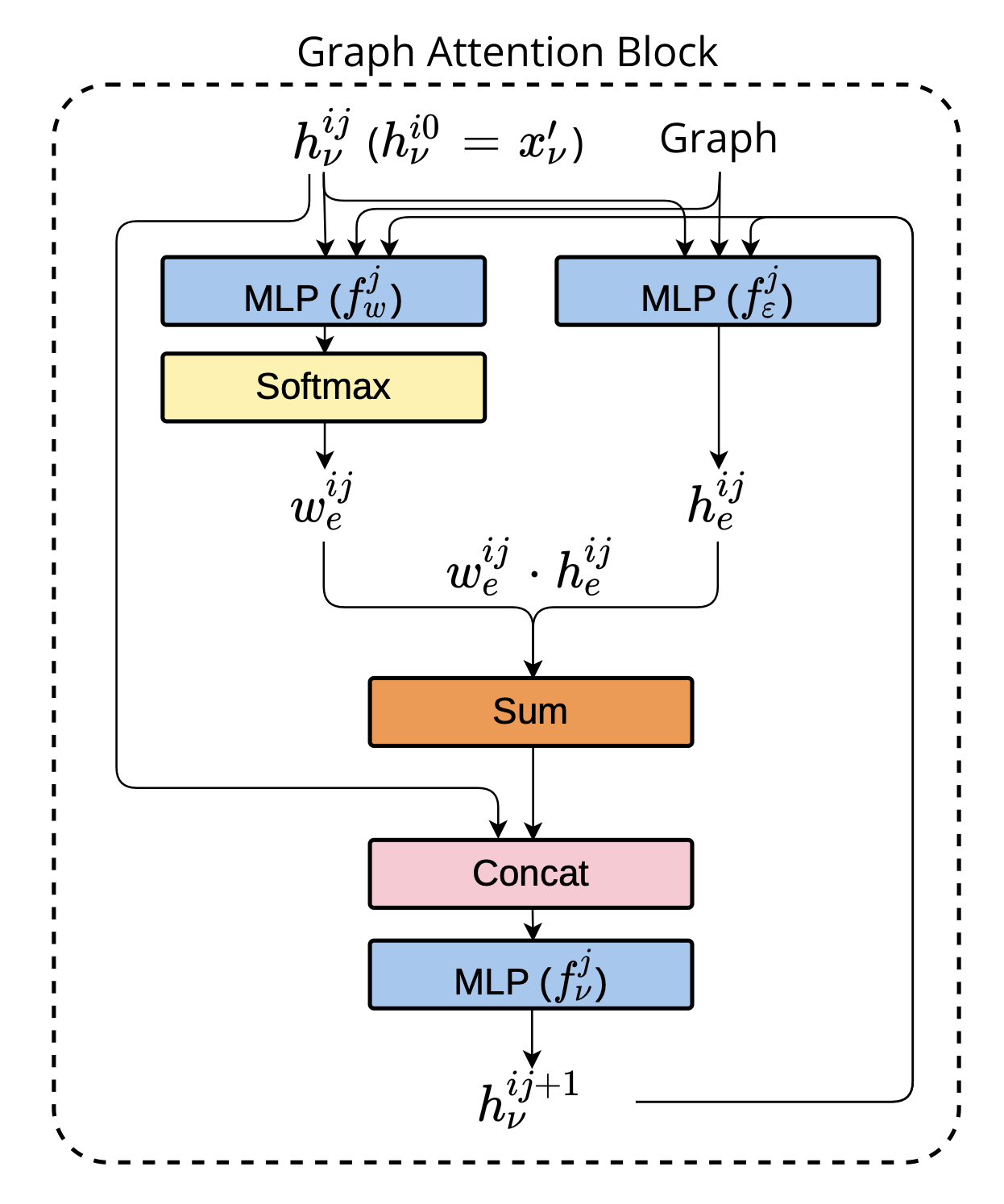}
\caption{The EggNet architecture. $i$ corresponds to each EggNet iteration, and $j$ corresponds to each message passing step. An EggNet iteration generally consists of a graph attention block and a KNN. The first iteration does not perform graph attention and the last iteration does not perform KNN.}
\label{fig:RGAT}
\end{center}
\vskip -0.2in
\end{figure*}

The EggNet architecture is illustrated in \cref{fig:RGAT}. Node features are first encoded with a multi-layer perception (MLP) node encoder $f_\mathrm{ENC}$. The encoded node features are then taken as inputs by a number of iterations where we first update the graph node embeddings and then the edges by connecting the K-Nearest Neighbors (KNN) in the node embedding space. In the first iteration ($i = 0$), each node is projected into a latent space $h_0$ by an MLP node network $f_0$. An MLP node decoder $f_\mathrm{DEC}$ then acts on $h_0$ and outputs a node embedding in $p_0$-space, followed by a KNN in $p_0$ which constructs the first graph $G_0$. In each of the next iterations ($i \geq 1$), multiple graph-attention-based \cite{veličković2018graph} message passing steps are performed to obtain the updated node representation $h_i$. The same node decoder $f_\mathrm{DEC}$ is used to obtain the updated node embedding in $p_i$-space. Similarly, KNN is performed in $p_i$ and updates the graph $G_i$.

In each message passing step, an edge representation ($e_{ij}$, where $i$ denotes the current EggNet iteration and $j$ denotes the current message passing step), and an edge weight $w_{ij}$ are learned from the two connecting nodes for each edge with the edge networks $f_\epsilon^j$ and $f_w^j$, respectively. The node representation $h_{ij}$ then is updated by a node network $f_\nu^j$ in each message passing step, which takes as the inputs the weighting sum of all connecting edges for each node. The $h_{ij}$ from the last message passing step ($h_{i(-1)}$) is then taken as $h_i$ that enters the node decoder $f_\mathrm{DEC}$.

The EggNet model is trained with the contrastive loss. For each edge connection, the individual loss term is given by:
\begin{equation}
    l = y d^2 + \left(1-y\right){\max}^2\left(0, m-d\right),
\end{equation}
where the first term is an attractive loss that brings together nodes belonging to the same particles, and the latter is a repulsive hinge loss that separates nodes belonging to different particles up to a given margin distance $m$. $y$ is 1 for an edge connecting spacepoints coming from the same particle (true edge), and 0 otherwise (fake edge). $d$ is the Euclidean distance between two nodes in the $p_j$-space in the last EggNet iteration ($p_{-1}$), and $m$ is set to 1 in our study. Due to the computation limit, we select representative edges to compute the total loss. The representative edges are pooled from three types, namely all true edges, randomly selected edges and all KNN edges. Since the majority of randomly selected edges are fake edges, the combination of true and randomly selected edges ensure that we optimize the model based on both true and fake edges. KNN edges represent the most true-like edges (edges connecting the nodes that are the nearest). Sampling from these edges allow us to emphasize the fake edges that look like true edges.

In the clustering step, $p_{-1}$ is taken as inputs of DBSCAN and a track label is output for each node. Two essential parameters for DBSCAN are the radius $\epsilon$ and the minimum number of nodes \textsc{min\_samples}. DBSCAN is only run for the inference and is not used in the training.

\section{Segmented Graph Training}

Due to the multiple iterations in the EggNet architecture, the computational cost, including computing time and memory, can be high. To mitigate the computational challenge, we train the EggNet with a subset of spacepoints for each training step, which significantly reduces the graph size during the training. In our study, each subset is selected randomly by a window of the azimuthal angle $\phi$. By randomly selecting the $\phi$ window, we prevent the model from being biased by certain $\phi$ range. While the size of the $\phi$ window is fixed initially, we allow the size to grow when no enough spacepoints are selected (the requirement is $> k$ in order to perform KNN search), and shrink when too many spacepoints are selected given the criteria. For each case, we perform a binary search of the window size, with 0 as the minimal window size and $2 \pi$ as the maximal window size. The initial window size is a hyperparameter, which has to be small enough to sufficiently reduce the computational cost, while large enough to retain the completeness of particle tracks.

In the following, the initial window size is set to $\Delta\phi = \frac{\pi}{5}$, and no maximal number of spacepoints is set. The segmentation is only applied during the training and not in the inference. Although it is possible to perform segmentation for the inference, we leave this task for future work.

\section{Evaluation on TrackML Dataset}

In the following, we test the EggNet tracking pipeline with the full TrackML dataset, which is a simulation of a generic tracking detector under the HL-LHC pileup condition. Each event contains an average of 9K particles and 108K spacepoints. Data are split to 2,000 events for training, and 400 for testing. We train the EggNet models with 12 input node features $\nu \in \mathbb{R}^{12}$ (described in \cite{ExaTrkX:2021abe}). All MLPs in the EggNet models consist of 3 hidden layers, each with a width of 128 neurons. Each intermediate layer in these networks uses a SiLU \cite{ELFWING20183} activation function and layer normalization \cite{ba2016layer}. The last layer of $f_\mathrm{DEC}$ uses a Tanh activation function, and the output is further normalized $p^\prime = \frac{p}{|p|}$. We use a dimension of 128 for all the $h$-space, and 24 for the $p$-space. Each EggNet model consists of 5 EggNet iterations ($i \leq 4)$, and each graph attention block consists of 8 message passing steps ($j \leq 7$). We consider $k=10$ for all KNN steps in the training. For DBSCAN, we find $\epsilon = 0.1$ to be an optimal value and set \textsc{min\_samples} to 3.

EggNet models are implemented with PyTorch \cite{NEURIPS2019_9015} and PyTorch Lightning \cite{Falcon_PyTorch_Lightning_2019} in the ACORN framework \cite{acorn} and optimized with Adam \cite{adam} with a learning rate of $2 \times 10^{-4}$. The training uses a batch size of 1 event and is performed for $\sim$1000 epochs. We employ Rapids cuML \cite{rapids} for KNN and DBSCAN, which enables GPU computation for these algorithms and significantly speeds up the process compared to CPU computation. For the following, the whole EggNet tracking pipeline is run on an NVIDIA A100 GPU.

We evaluate the track performance on all track candidates extracted from the DBSCAN clusters. A track candidate is matched to a particle if more than 50\% of the track spacepoints come from that particle. Three metrics are defined as follows:
\begin{itemize}
    \item Efficiency: The fraction of particles to which at least 1 track candidate is matched.
    \item Duplication rate: The fraction of track candidates that are matched to the same particles.
    \item Fake rate: The fraction of track candidates that are not matched to any particle.
\end{itemize}

\begin{figure}[htbp]
\begin{center}
\includegraphics[width=0.6\columnwidth]{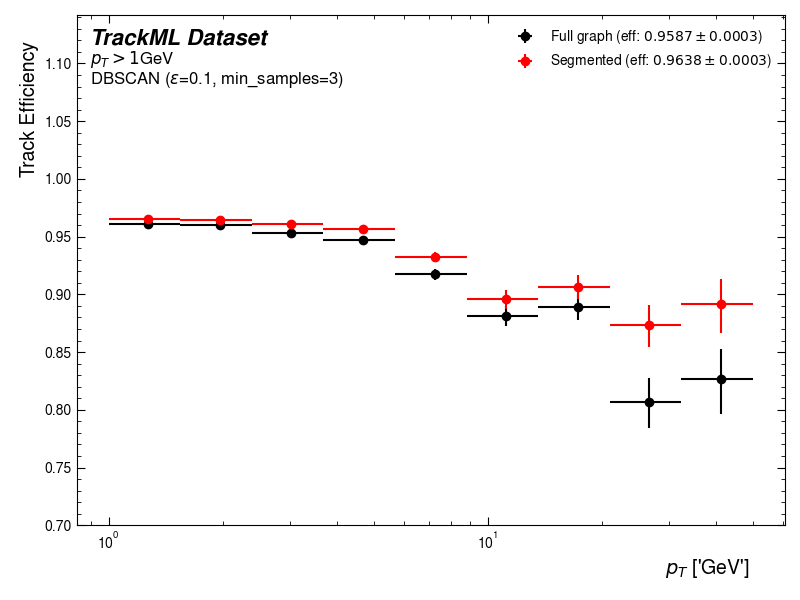}
\caption{Track efficiency as a function of particle transverse momentum ($p_\mathrm{T}$), shown for both full graph and segmented graph training. About 467K particles are considered for the evaluation.}
\label{fig:track_eff}
\end{center}
\end{figure}

As shown in \cref{fig:track_eff}, we plot the track efficiency across different particle transverse momentum ($p_\mathrm{T}$) bins. Performance is shown for both the full graph training and the segmented graph training. About 467K particles are considered for the evaluation. We observe that both training strategies perform very similarly, giving an efficiency of $0.9587 \pm 0.0003$ ($0.9638 \pm 0.0003$), a duplication rate of $0.0276 \pm 0.0002$ ($0.0428 \pm 0.0003$), and a fake rate of $0.0019 \pm 0.0001$ ($0.0021 \pm 0.0001$) for the full graph training (segmented graph training). Both cases yield a higher track efficiency in lower $p_\mathrm{T}$, and lower efficiency in higher $p_\mathrm{T}$, which, as also observed in the case of the reduced TrackML dataset \cite{Calafiura:2024qhv}, is likely due to imbalanced training data statistics across $p_\mathrm{T}$.

\begin{figure*}[htbp]
\begin{center}
\subfloat[][]{\includegraphics[width=0.499\columnwidth]{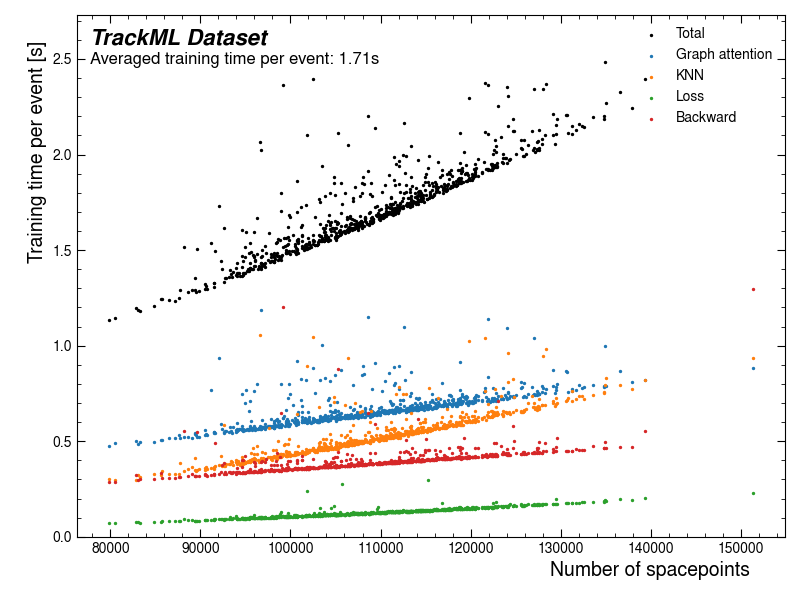}\label{fig:training_time}}
\subfloat[][]{\includegraphics[width=0.499\columnwidth]{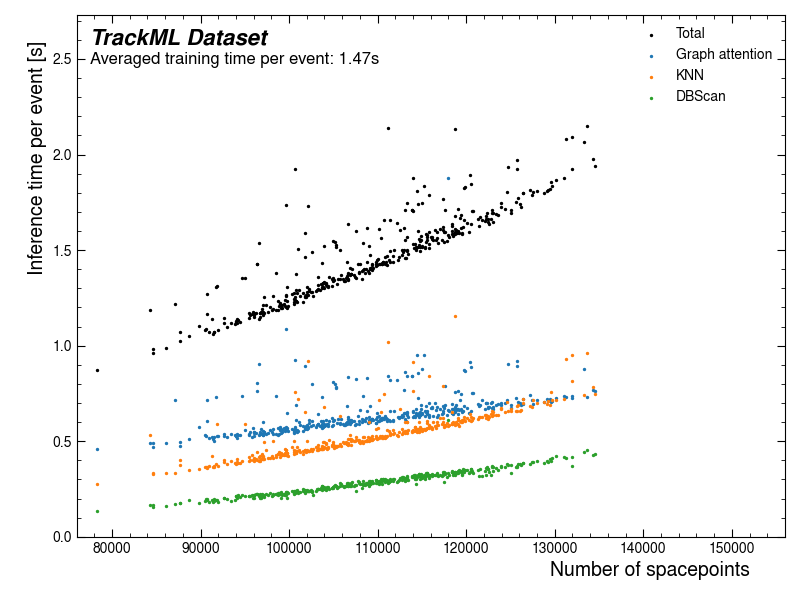}\label{fig:inference_time}}
\caption{Training (a) and inference (b) time for each event versus the number of spacepoints. The computing time is also broken down different components, including Graph Attention, KNN, calculation of loss, backward propagation and DBSCAN.}
\label{fig:computing_time}
\end{center}
\end{figure*}

In \cref{fig:computing_time}, we show the full graph training and inference time versus the total number of spacepoints for each event. The averaged total computing time per event is around 1.71 seconds for training and 1.47 seconds for inference. The total computing time is also broken down into different components, including Graph Attention, KNN, calculation of loss, backward propagation and DBSCAN. The majority of the computing time comes from graph attention and KNN. Particularly, KNN scales quadratically with the event size, which can become challenging when the event size is further increased. Our future work will focus on replacing the KNN algorithm with an alternative method or reducing the input size for KNN in order to improve the scalability.

\begin{figure*}[htbp]
\begin{center}
\subfloat[][]{\includegraphics[width=0.499\columnwidth]{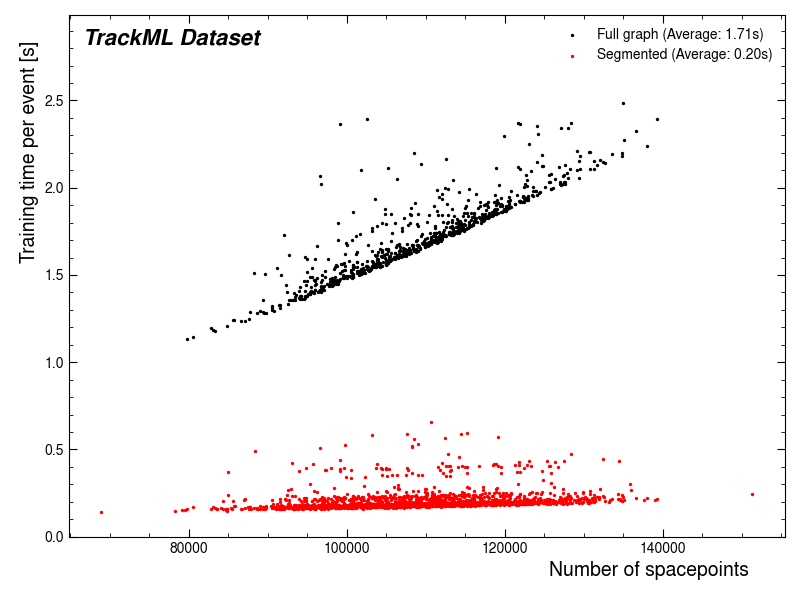}\label{fig:training_time_seg}}
\subfloat[][]{\includegraphics[width=0.499\columnwidth]{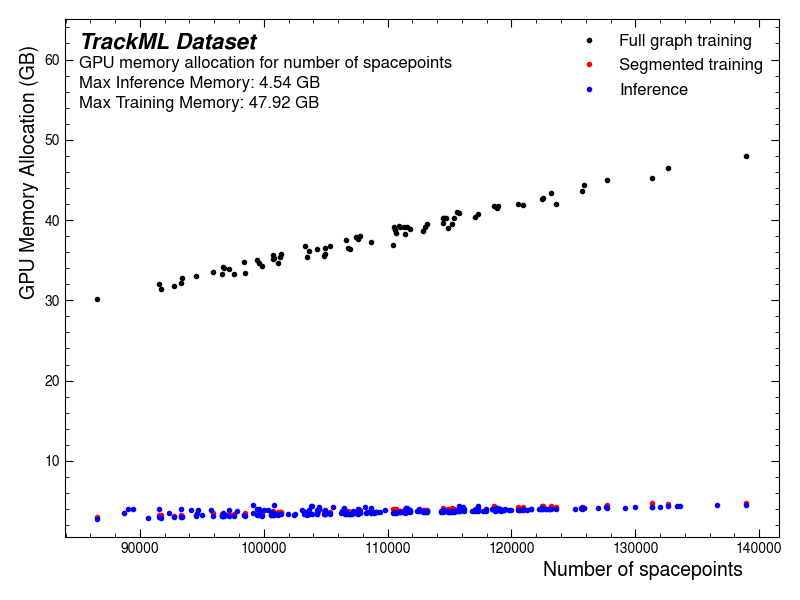}\label{fig:memory_usage_seg}}
\caption{(a) Comparison of the training time between full graph and segmented graph training. (b) GPU memory requirement for full graph training, segmented graph training, and inference.}
% \label{fig:training_time_seg}
\end{center}
\end{figure*}

A comparison of the training time between full graph and segmented graph training is also shown in \cref{fig:training_time_seg}. Training the EggNet with the segmented graphs (subsets of spacepoints) significantly reduces the training time by about a factor of 10, and at the same time has minimal impacts on the physics performance as shown previously.

Finally, we benchmark the required GPU memory for training and inference, as a function of the number of spacepoints, and compare between full graph and segmented graph training, as shown in \cref{fig:memory_usage_seg}. Despite the high demand of GPU memory in the full graph training, segmented graph training reduces the GPU memory requirement by about a factor of 10.

\section{Conclusions}

In this paper, we evaluated the EggNet physics and computing performance on the full TrackML dataset. To reduce the computational cost during the training, we introduced the strategy of segmented graph training. Training on segmented graphs significantly reduces the training time and memory requirement during the training, and at the same time is able to provide comparable track performance to the full graph training. For future work, we will focus on further improving the computing performance. In particular, we aim to improve the scalability of KNN, and apply graph segmentation also to the inference. Furthermore, we plan to evaluate the EggNet performance on a more realistic dataset such as ATLAS ITk simulation.

\section*{Acknowledgements}

% JC is supported by the Scientific Discovery through Advanced Computing (SciDAC) program funded by U.S. Department of Energy, Office of Science, Advanced Scientific Computing Research and High Energy Physics.

This work was supported by the DOE HEP Center for Computational Excellence at Lawrence Berkeley National Laboratory under B\&R KA2401045.
%
% BibTeX or Biber users please use (the style is already called in the class, ensure that the "woc.bst" style is in your local directory)
\bibliography{mybib} % Replace "your_bib_file" with the actual name of your .bib file
% \begin{thebibliography}{}
% %
% % and use \bibitem to create references.
% % Format for Journal Reference
% Journal Author, Article title. Journal \textbf{Volume}, page numbers (year). \url{https://doi.org/Article-DOI-number}
% Book Author, \textit{Book title} (Publisher, place, year) page numbers
% \end{thebibliography}

\end{document}